\documentstyle[12pt,preprint,aps,floats]{revtex}
\def\NPB#1#2#3{Nucl. Phys. {\bf B#1}, #3 (19#2)}

\def\PLB#1#2#3{Phys. Lett. {\bf B#1}, #3 (19#2)}

\def\PRD#1#2#3{Phys. Rev. {\bf D#1}, #3 (19#2)}
\def\PRL#1#2#3{Phys. Rev. Lett. {\bf#1}, #3 (19#2)}

\def\ZPC#1#2#3{Zeit. f\"ur Physik {\bf C#1}, #3 (19#2)}

\def\JHEP#1#2#3{JHEP {\bf #1}, #3 (19#2)}

\input epsf.tex
\newcommand{\postscript}[2]{\setlength{\epsfxsize}{#2\hsize}
   \centerline{\epsfbox{#1}}}

\newcommand{\talpha}{\tilde{\alpha}}
\newcommand{\tr}{\text{ tr}}
\newcommand{\bigH}{{\cal H}}
\newcommand{\mheavy}{m_{\rm heavy}}
\newcommand{\mlight}{m_{\rm light}}
\newcommand{\boldm}{\mbox{\boldmath $m$}}
\newcommand{\bolde}{\mbox{\boldmath $e$}}
\newcommand{\boldN}{\mbox{\boldmath $N$}}

\newcommand{\tev}{\text{ TeV}}
\newcommand{\gev}{\text{ GeV}}

\tighten

\begin{document}

\preprint{
\noindent
\begin{minipage}[t]{3in}
\begin{flushleft}
May 1999 \\
\end{flushleft}
\end{minipage}
\hfill
\begin{minipage}[t]{3in}
\begin{flushright}
IASSNS--HEP--99--13\\
RU--99--05\\
hep-ph/9905292\\
\vspace*{.7in}
\end{flushright}
\end{minipage}
}

\title{Naturally Heavy Scalars in \\
Supersymmetric Grand Unified Theories }

\author{Jonathan Bagger,$^{ab}$ Jonathan L. Feng,$^b$ Nir Polonsky$^c$
\vspace*{.2in}
}
\address{${}^{a}$Department of Physics and Astronomy,
Johns Hopkins University\\
Baltimore, MD 21218  USA}
\address{${}^{b}$School of Natural Sciences,
Institute for Advanced Study\\ Princeton, NJ 08540 USA}
\address{${}^{c}$Department of Physics and Astronomy,
Rutgers University\\ Piscataway, NJ 08854 USA
\vspace*{.2in}
}

\maketitle

\begin{abstract}
The supersymmetric flavor, $CP$ and Polonyi problems are hints that
the fundamental scale of the soft supersymmetry breaking parameters
may be above a TeV, in apparent conflict with naturalness.  We
consider the possibility that multi-TeV scalar masses are generated by
Planck- or unification-scale physics, and find the conditions under
which the masses of scalars with large Yukawa couplings are driven,
radiatively and asymptotically, to the weak scale through
renormalization group evolution.  Light third generation scalars then
satisfy naturalness, while first and second generation scalars remain
heavy to satisfy experimental constraints.  We find that this
mechanism is beautifully realized in the context of grand unified
theories. In particular, the existence of right-handed neutrinos plays
an important role in allowing remarkably simple scenarios.  For
example, for SO(10) boundary conditions with the squared masses of
Higgs scalars double those of sleptons and squarks, we find that the
entire scalar mass scale may be increased to 4 TeV at the unification
scale without sacrificing naturalness.
\end{abstract}



\section{Introduction}
\label{sec:introduction}

The Achilles heel of supersymmetry lies, arguably, in the difficulty
of satisfying naturalness while simultaneously decoupling
supersymmetric effects from low-energy experimental observables.
Naturalness is typically taken to require supersymmetric particle
masses below $\sqrt{(4\pi/\alpha)} M_W \sim 1 \tev$, a few times the
weak scale.  In contrast, without an understanding of small scalar
mass splittings, mixing angles, and $CP$-violating phases, a variety
of experimental measurements suggest that supersymmetric scalar masses
must lie well above a TeV.

The experimental constraints have varying degrees of significance and
model dependence.  Examples include the following:

\begin{itemize}

\item The kaon system.  Constraints from the $K$ system depend on the
flavor and chirality structure of the scalar quarks.  Roughly
speaking, however, for moderate degeneracies, the constraint from
$\epsilon_K$ is most severe; it requires squark masses of order 100
TeV.  Likewise, the $K-\bar{K}$ mass splitting needs squark masses of
order 10 TeV.  (See Ref.~\cite{constraints} for details.)

\item Electric dipole moments.  The electric dipole moments of the
neutron and electron typically require multi-TeV squark and slepton
masses if the relevant phases are not suppressed~\cite{edm}.  EDMs are
flavor-conserving and so cannot be suppressed by scalar degeneracy;
for example, they imply severe constraints on scalar masses (or
phases) even in models with gauge-mediated supersymmetry
breaking.~\cite{GMCP}.

\item The proton lifetime.  Lower bounds on the proton lifetime place
strong constraints on supersymmetric grand unified theories (GUTs),
especially for large $\tan\beta$~\cite{proton}.  The predicted decay
rates depend on the (super)heavy and the light supersymmetric particle
spectra; they are greatly suppressed for multi-TeV scalar masses.

\item The Polonyi problem. Many supersymmetric theories contain late
decaying scalar fields that ruin the successful predictions of Big
Bang nucleosynthesis~\cite{Polonyi}.  This problem is independent of
scalar degeneracies or phases, but may be solved if the scale of
supersymmetry breaking is of order 10 TeV~\cite{moroi}.
\end{itemize}

In addition, many models predict scalar masses hierarchically larger
than gaugino masses~\cite{anomaly}.  Given the current bounds on
gaugino masses, scalar masses must then be well above a TeV.

The conflict between naturalness and experimental constraints may be
resolved by observing that, roughly speaking, naturalness restricts
the masses of scalars with large Yukawa couplings, while experiment
constrains the masses of scalars with small Yukawas~\cite{dgpt}.
Naturalness affects particles that are strongly coupled to the Higgs
sector, while experimental constraints are strongest in sectors with
light fermions that are plentifully produced.  This suggests that
naturalness and experimental constraints may be simultaneously
satisfied by an ``inverted hierarchy'' approach, in which light
fermions have heavy superpartners, and vice versa.  In particular,
third generation scalars with masses $\mlight \alt 1 \tev$ satisfy
naturalness constraints, while first and second generation scalars at
some much higher scale $\mheavy$ avoid many experimental difficulties.
A number of possibilities have been proposed to dynamically generate
scalar masses at two hierarchically separated
scales~\cite{dp,U1,horizontal}.

In this paper, we shall investigate a new mechanism for generating an
inverted hierarchy.  We will assume that all scalar masses are of
order $\mheavy$ when produced at some scale, say the GUT or Planck
scale.  We will then use renormalization group evolution to create an
inverted hierarchy at the weak scale.  This mechanism works because
the renormalization group equations automatically suppress the masses
of scalars associated with fermions which have large Yukawa couplings.
The masses of these scalars (and only these scalars) are driven to
$\mlight$.  This mechanism was investigated recently in models with
large top and bottom Yukawa couplings (and all others
small)~\cite{fkp}.  It was shown that, for particular high scale
boundary conditions, all scalars with large Yukawas may be driven
asymptotically to $\mlight$ in the infrared.  Stops, sbottoms, and
Higgs scalars are then naturally light at the weak scale, while the
masses of all other scalars remain heavy.  In this way a very large
hierarchy can be created dynamically.

In what follows we will consider the well-motivated supersymmetric
GUTs, with special emphasis on SO(10) unified theories.  As is
well-known, such theories, among other merits, successfully predict
gauge coupling unification and naturally accommodate massive
neutrinos.  In the present context, however, they have even more
virtues.  First, they naturally provide the universal and large third
generation Yukawa couplings required to implement the inverted
hierarchy.  Second, as we will see, the existence of a right-handed
neutrino plays an important part in creating a plausible fixed point
structure.  These virtues make grand unified theories the natural
setting for the radiatively generated inverted hierarchy mechanism.

We will begin in Sec.~\ref{sec:FP} with a discussion of the fixed
point structure of the scalar mass renormalization group equations
(RGEs).  In Sec.~\ref{sec:mssm} we will analyze the inverted hierarchy
in grand unified models with right-handed neutrinos.  Finally, in
Sec.~\ref{sec:guts} we will consider inverted hierarchies generated in
supersymmetric GUTs with evolution between the Planck and GUT scales.
In each case we illustrate the size of the hierarchies that may be
achieved using the scalar mass fixed-point framework.  We shall see
that the scalar masses may be pushed as high as $\sim 4$ TeV.  Such a
large hierarchy significantly ameliorates most experimental
difficulties.  More detailed conclusions, as well as a discussion of
implications for low energy experiments and high energy colliders, are
contained in Sec.~\ref{sec:conclusions}.

In the Appendix we collect the RGEs used in this analysis.  For models
above the GUT scale, these RGEs include, for the first time, the
two-loop pure Yukawa terms.  They also correct some one-loop results
previously given in the literature.

\section{Scalar mass fixed points}
\label{sec:FP}

Supersymmetric theories are natural if the conditions for electroweak
symmetry breaking,
\begin{eqnarray}
\frac{1}{2} m_Z^2 &=& \frac{m_{H_d}^2 - m_{H_u}^2 \tan^2\beta }
{\tan^2\beta -1} - \mu^2, \nonumber \\
2 m_3^2 &=& (m_{H_u}^2 + m_{H_d}^2 + 2 \mu^2)\sin2\beta \ ,
\label{finetune}
\end{eqnarray}
are free of large cancellations.\footnote{The parameters $m_{H_u}$ and
$m_{H_d}$ are the soft supersymmetry breaking Higgs boson masses,
$m_3^2$ is the soft bilinear scalar coupling of the two Higgs
doublets, $\mu$ is the Higgsino mass parameter, and $\tan\beta =
\langle H_u^0 \rangle /\langle H_d^0\rangle$ is the usual ratio of
Higgs vacuum expectation values.}  These conditions apply to the
RGE-improved effective potential, so the supersymmetric parameters of
Eq.~(\ref{finetune}) must be evaluated at the weak scale.  {\em A
priori}, naturalness bounds only the parameters entering
Eq.~(\ref{finetune}), and, for example, even the requirement on
$m_{H_d}$ is relaxed for large $\tan\beta$.  Nevertheless, naturalness
affects all other supersymmetric parameters to the extent that they
renormalize these relations.

For all the reasons given in the previous section, we are motivated to
consider theories in which the scalar masses are of order $\mheavy$ at
some high scale $M_0$, which we take to be the GUT or Planck scale.
For arbitrary initial conditions, it is clear that the scalar masses,
including $m_{H_u}$ and $m_{H_d}$, are of order $\mheavy$ at all
scales, and the theory must be fine-tuned.  However, if the RGEs have
(approximate) infrared fixed points where these scalar masses vanish,
appropriate boundary conditions will lead to weak-scale masses that
are small and insensitive to $\mheavy$, thereby preserving
naturalness.

We must therefore identify the infrared fixed points of the scalar
mass renormalization group equations.  The RGEs for supersymmetric
theories are well-known (up to two-loops), but complicated~\cite{rge}.
However, in the scenarios we are considering, we can make several
simplifications and easily extract the essential features of the RGEs.

In a general supersymmetric theory, the one-loop RGEs for scalar
squared masses $m^2$ are schematically of the form
\begin{equation}
\dot{m}^2 \sim -Y m^2 + \talpha M_{1/2}^2 - Y A^2 \ ,
\label{scalarRGE}
\end{equation}
where $M_{1/2}$ and $A$ represent gaugino masses and trilinear scalar
couplings, respectively.\footnote{We have omitted a hypercharge trace
term which vanishes for supersymmetric GUTs.}  Here, and in the rest
of the paper, we use the following notation and conventions: $t \equiv
\ln \left(M_0^2/Q^2\right)$, $\dot{(\ )} \equiv \frac{d}{dt}$,
$\talpha \equiv g^2/16 \pi^2 = \alpha/4 \pi$, and $Y \equiv h^2/16
\pi^2$, where $g$ and $h$ are gauge and Yukawa couplings,
respectively.  For each term in Eq.~(\ref{scalarRGE}), overall signs
are determined as shown, but (positive) numerical coefficients are
suppressed.

{}From Eq.~(\ref{scalarRGE}) we see that large $M_{1/2}$ and $A$
parameters typically destabilize light scalar masses.  Therefore we
require\footnote{Gaugino masses satisfy $\frac{d}{dt}\left(
M_{1/2}/\talpha \right) = 0$ at one-loop; they cannot be driven from
$\mheavy$ at $M_0$ to $\mlight$ in the infrared.  For $A$-parameters,
the story is more complicated.  We will simply assume $A \sim
M_{1/2}\sim \mlight$ at all scales.}  $M_{1/2}, A \sim \mlight$ at the
scale $M_0$.  Note that the hierarchy $M_{1/2}, A \ll m$ at $M_0$ is
natural and may be generated in a number of ways, for example, by an
approximate $R$-symmetry, or through mechanisms that differentiate
dimension-one and dimension-two soft supersymmetry breaking
parameters.  Examples are given in Refs.~\cite{fkp,ftp}.

Let us also assume that all large Yukawa couplings are unified at
$M_0$.  Then, if we ignore differences in the Yukawa evolution, we
find that the RGEs for the scalar masses take the simple form
\begin{equation}
\dot{\boldm}_i^2 = - Y \boldN_{ij} \boldm_j^2 \ ,
\end{equation}
where the subscripts run over all scalar fields in the theory, and
$\boldN$ is a matrix of positive constants determined by color and
SU(2) factors.  This set of equations is easily solved by decomposing
arbitrary initial conditions into components parallel to the
eigenvectors of $\boldN$, each of which then evolve independently.
Indeed, if $\boldN$ has eigenvectors $\bolde_i$ with eigenvalues
$\lambda_i$, the initial conditions
\begin{equation}
\boldm^2(t=0) = \sum c_i \bolde_i
\end{equation}
evolve to
\begin{equation}
\boldm^2(t=t_f) = \sum c_i\bolde_i \, {\rm exp}
\left[ - \lambda_i \int_0^{t_f} Y dt\right] \ .
\label{mend}
\end{equation}
We see that eigenvectors with large positive eigenvalues are
asymptotically and rapidly crunched to zero.  For initial conditions
dominated by such eigenvectors, the final masses are greatly
suppressed relative to their initial values.

Of course, the size of the scalar mass hierarchy depends not only on
the initial conditions, but also on the evolution interval $t_f$ and
the initial value of the Yukawa coupling.  We expect a large hierarchy
for large $t_f$ and Yukawa couplings near their quasi-fixed
points~\cite{qfp}.  In the following sections, we determine
numerically the suppression factors, or crunch factors, that are
possible in supersymmetric GUTs both below and above the unification
scale.

\section{MSSM with Right-handed Neutrino}
\label{sec:mssm}

In this section, we consider a model with the particle content of the
minimal supersymmetric standard model (MSSM) with a right-handed
neutrino $N$.  We assume that soft supersymmetry breaking masses are
generated at the unification scale, as may be the case, for example,
in strongly-coupled string theories where the GUT scale is also the
string and, hence, the supergravity scale.

The superpotential is given by
\begin{equation}
W = h_t H_u Q U + h_b H_d Q D + h_{\tau} H_d L E + h_n H_u L N \ ,
\end{equation}
where the matter fields $Q$, $U$, $D$, $L$, $E$, and $N$ are those of
the third generation and all other Yukawa couplings may be neglected
for our analysis.  We also neglect off-diagonal scalar masses for the
moment; their effects and constraints will be discussed in
Sec.~\ref{sec:conclusions}.

For this model, omitting gaugino masses and $A$ terms as discussed in
Sec.~\ref{sec:FP}, the RGEs for scalar masses are
\begin{equation}
\dot{\boldm}^2 = - \left[
\begin{array}{cccccccc}
3Y_t + Y_n & 3Y_t & 3Y_t & 0 & 0 & Y_n & 0 & Y_n \\
2Y_t & 2Y_t & 2Y_t & 0 & 0 & 0 & 0 & 0 \\
Y_t & Y_t & Y_t+Y_b & Y_b & Y_b & 0 & 0 & 0 \\
0 & 0 & 2Y_b & 2Y_b & 2Y_b & 0 & 0 & 0 \\
0 & 0 & 3Y_b & 3Y_b & 3Y_b+Y_{\tau} & Y_{\tau} & Y_{\tau} & 0 \\
Y_n & 0 & 0 & 0 & Y_{\tau} & Y_{\tau}+Y_n & Y_{\tau} & Y_n \\
0 & 0 & 0 & 0 & 2Y_{\tau} & 2Y_{\tau} & 2Y_{\tau} & 0 \\
2Y_n & 0 & 0 & 0 & 0 & 2Y_n & 0 & 2Y_n \end{array} \right] \boldm^2 ,
\end{equation}
where ${\boldm^2} = (m_{H_u}^2, m_{U}^2, m_{Q}^2, m_{D}^2, m_{H_d}^2,
m_{L}^2, m_{E}^2, m_{N}^2)^T$.  At the GUT scale $M_G$, the Yukawa
couplings are unified with $Y = Y_t = Y_b = Y_{\tau} = Y_n$.  If we
assume, for the moment, that they remain approximately degenerate at
lower scales, the evolution equations simplify to
\begin{equation}
\dot{\boldm}^2 = - Y {\boldN} \boldm^2 \ ,
\end{equation}
where
\begin{equation}
{\boldN}= \left[
\begin{array}{cccccccc}
4 & 3 & 3 & 0 & 0 & 1 & 0 & 1 \\
2 & 2 & 2 & 0 & 0 & 1 & 0 & 1 \\
1 & 1 & 2 & 1 & 1 & 0 & 0 & 0 \\
0 & 0 & 2 & 2 & 2 & 0 & 0 & 0 \\
0 & 0 & 3 & 3 & 4 & 1 & 1 & 0 \\
1 & 0 & 0 & 0 & 1 & 2 & 1 & 1 \\
0 & 0 & 0 & 0 & 2 & 2 & 2 & 0 \\
2 & 0 & 0 & 0 & 0 & 2 & 0 & 2 \end{array} \right] \ .
\end{equation}
The eigenvectors of $\boldN$, and their associated eigenvalues
$\lambda_i$, are
\begin{eqnarray}
\bolde_1 : && 8, (2,1,1,1,2,1,1,1) \nonumber \\
\bolde_2 : && 6, (2,1,0,-1,-2,0,-1,1) \nonumber \\
\bolde_3 : && 4, (0,1,1,1,0,-3,-3,-3) \nonumber \\
\bolde_4 : && 2, (0,-1,0,1,0,0,-3,3) \ ,
\end{eqnarray}
along with four eigenvectors with zero eigenvalue.  Components of the
initial conditions along $\bolde_1$, and to a lesser extent
$\bolde_2$, $\bolde_3$, and $\bolde_4$, are rapidly suppressed during
renormalization group evolution by factors of ${\rm exp} [ -\lambda_i
\int Y dt]$.

Note that the boundary condition given by $\bolde_1$,
\begin{equation}
m_{U}^2= m_{Q}^2= m_{D}^2= m_{L}^2= m_{E}^2= m_{N}^2 = \frac{1}{2}
m_{H_u}^2 = \frac{1}{2} m_{H_d}^2 \ ,
\label{mssmbc}
\end{equation}
is remarkably simple, and is consistent with minimal SO(10)
unification, in which all matter fields arise from a single {\boldmath
$16$} multiplet, both Higgs fields are contained in a single
{\boldmath $10$} multiplet, and the GUT breaking $\text{SO(10)} \to
\text{SU(3)} \times \text{SU(2)} \times \text{U(1)}$ takes place in
one step.\footnote{Of course, such a minimal unification scenario is
not consistent with the observed light fermion spectrum and CKM
mixing. However, these issues may be resolved, for example, by
non-renormalizable couplings, which are irrelevant to our analysis of
the large ${\cal{O}}(1)$ Yukawa couplings. The minimal model may
therefore be used to illustrate our results.}  This simplicity is {\it
not} automatic.  For example, equivalent analyses without a
right-handed neutrino yield eigenvectors that are far from simple, and
moreover, are not compatible with GUT unification.  The right-handed
neutrino plays a critical role in leading us to a plausibly simple
boundary condition.

In reaching the above conclusions, we have made a number of
approximations that must be examined.  First, we have neglected the
fact that the Yukawa couplings evolve independently, and, in
particular, the fact that the leptonic and hadronic Yukawas differ
significantly at lower energy scales.  Second, we have not taken into
account the decoupling of the right-handed neutrino $N$ at some scale.
Both of these effects imply that the above eigenvectors do not remain
eigenvectors during the full renormalization group evolution.
Finally, as we will see below, there may also be corrections from
numerically significant two-loop terms.

To determine the importance of these effects, we now solve the RGEs
numerically.  It is, of course, possible to include all terms up to
two-loops in the numerical analysis.  However, doing so requires
specifying many additional parameters, and the analysis becomes highly
model-dependent.  Fortunately, many terms have only a small effect on
the overall fixed point behavior and therefore may be neglected.  In
what follows, we will systematically omit all terms of order
$\mlight^2$ in the $m^2$ RGEs; this removes all dependence on the
unknown gaugino mass and $A$ parameters.  For consistency, we also
omit all other terms of similar size, and keep only the one-loop and
leading two-loop terms. The resulting RGEs are schematically of the
form
\begin{eqnarray}
\dot{\talpha} &=& b\talpha^2 + \talpha^2 Y \nonumber \\
\dot{Y} &=& -Y^2 + \talpha Y + Y^3 \label{mssmRGE} \\
\dot{m}^2 &=& -Y m^2 + Y^2 m^2 \ , \nonumber
\end{eqnarray}
where $b$ is the one-loop $\beta$-function coefficient.\footnote{Note
that $\talpha^2 \mheavy^2$ terms have been omitted in the $m^2$ RGE.
However, given the hierarchies we are able to achieve (see below),
these terms are still only of order $\mlight^2$.}  The complete RGEs
up to this order are presented in the Appendix.

The approximation used is formally equivalent to keeping the leading
and next-to-leading order terms in an expansion in ${\cal
O}(\sqrt{\talpha}, Y)$.  This expansion is motivated by the fact that,
at the unification scale, Yukawa couplings may be very large near
their quasi-fixed points.  For example, for $h \sim 2$, we have $Y
\sim 1/40$, while $\talpha \approx 1/300$.  The dominant two-loop
terms are therefore those of Eq.~(\ref{mssmRGE}).  We will see that
these terms may be significant.  On the other hand, while possibly
significant, they should not be too large: large two-loop effects
signal a breakdown of perturbativity, and our analysis is unreliable
in these regions. We will comment on such parameter regions below.

\begin{figure}[tbh]
\postscript{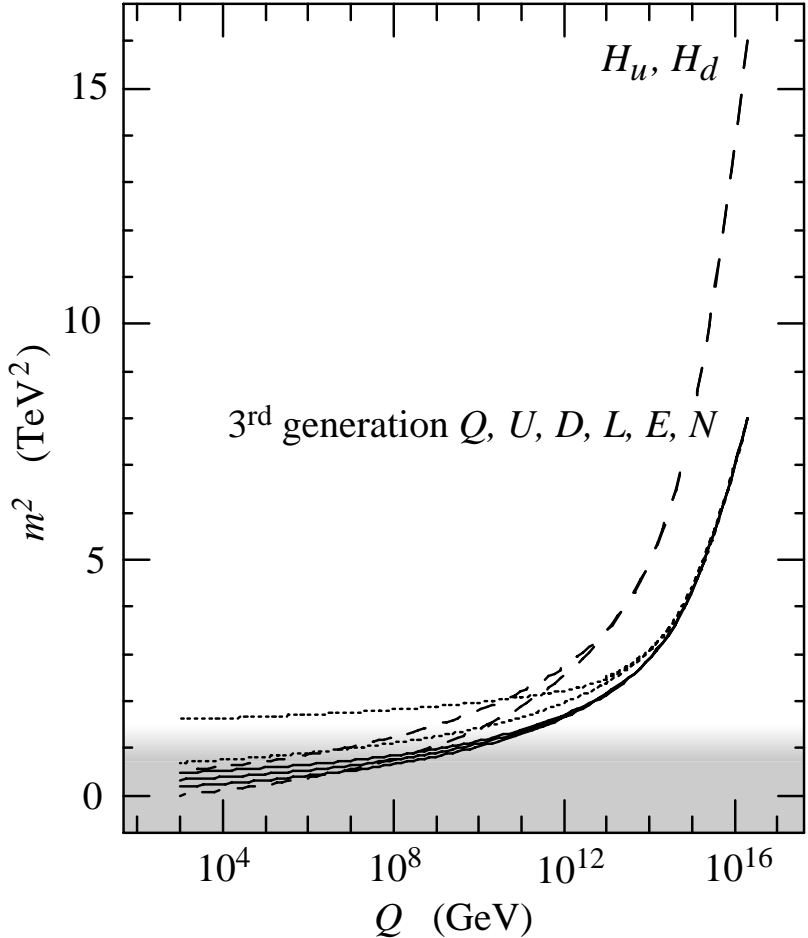}{0.65}
\caption{The renormalization group evolution of the Higgs (dashed) and
third generation squark (solid) and slepton (dotted) squared masses in
the MSSM with a right-handed neutrino for the boundary conditions of
Eq.~(\ref{mssmbc}) with $h_G = 2$ and $m_N = 10^{13}\gev$. First and
second generation scalar masses (not shown) are approximately
renormalization group invariant.  At the weak scale, $m_L = 1270\gev$,
$m_E = 830\gev$, $m_{H_u} = 710\gev$, $m_D = 690\gev$, $m_Q =
570\gev$, $m_U = 420\gev$, and $m_{H_d} = 50\gev$.  Note, however,
that neglected effects of order $\mlight^2$ modify solutions in the
shaded region. The suppression factor for this case is $S=20$ (see
text).}
\label{fig:mssmRGE}
\end{figure}

With these approximations, the theory is completely specified by two
parameters, $h_G$, the universal third generation Yukawa coupling at
the GUT scale, and $m_N$, the right-handed neutrino mass.  In
Fig.~\ref{fig:mssmRGE}, we show the renormalization group evolution of
the scalar mass parameters from the GUT scale $M_G \simeq 2\times
10^{16}$ GeV to the weak scale, for $h_G = 2$ and $m_N = 10^{13}\gev$.
We find that large scalar mass suppressions are possible.  Indeed,
even $\mheavy \sim 4 \tev$ can give rise to third generation scalars
with masses $\mlight \alt 1 \tev$.  Note also that the hierarchy is
created rapidly, in the first few decades of evolution.  This is the
region in which the gauge interactions, and hence, the Yukawa
couplings, remain roughly universal.  Below this scale, the Yukawa
couplings split substantially, but by then, the hierarchy is already
created and cannot be destroyed.  The assumption of universal Yukawa
couplings throughout the evolution interval is therefore a convenient
fiction.

In our analysis, we have neglected effects of order $\mlight$, which
clearly have little impact on the overall suppression of the scalar
mass scale.  They do, however, modify the RGEs at the weak scale when
the scalar masses are of order $\mlight$.  By omitting such effects,
we forfeit the possibility of investigating a number of topics,
including the details of electroweak symmetry breaking, as well as
quantitative determinations of finite corrections to fermion masses
and the residual fine-tuning required to keep the $Z$ mass light.
Here, we note only that our scenario shares the problem of proper
electroweak symmetry breaking generic to all supersymmetric scenarios
with large $\tan\beta$~\cite{largetgb}.  Several possible solutions
have been discussed in the literature.  For example, if the GUT group
is broken not in one step, but in a chain beginning with
$\text{SO(10)} \to \text{SU(5)} \times \text{U(1)}$, the resulting
U(1) $D$-terms may help induce proper electroweak symmetry
breaking~\cite{Dterms}.  These terms are parametrically of order
$\mheavy^2$, but are suppressed by the large $\text{U(1)}_{X}$ charge
of the singlet.  More generally, another source for
${\cal{O}}(\mlight^2)$ (but not ${\cal{O}}(\mheavy^2)$) perturbations
could be assumed.

\begin{figure}[t]
\postscript{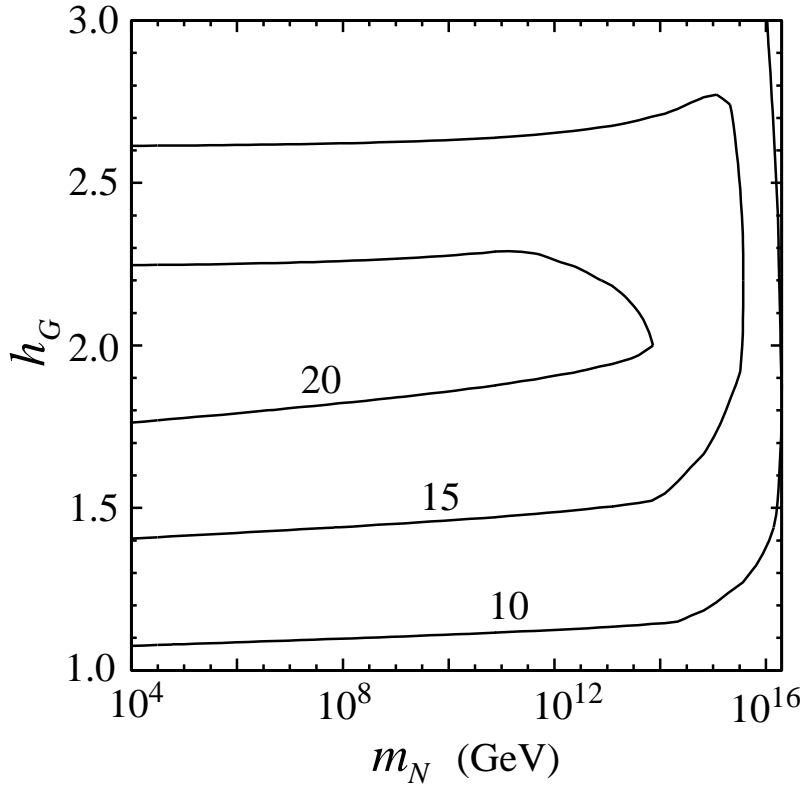}{0.65}
\caption{The suppression factor $S$ defined in the text for the MSSM
with a right-handed neutrino $N$ with initial boundary conditions
given by Eq.~(\ref{mssmbc}).  The parameter $m_N$ is the scale at
which the right-handed neutrino decouples, and $h_G$ is the value of
the universal Yukawa coupling at the GUT scale $M_G \simeq 2 \times
10^{16}$.}
\label{fig:mssmS}
\end{figure}

The large radiatively-generated hierarchy evident in
Fig.~\ref{fig:mssmRGE} exists for a wide range of parameters $m_N$ and
$h_G$.  To quantify the hierarchy, we characterize the scale of a
given scalar spectrum at renormalization scale $Q$ by the quantity
\begin{equation}
\bar{m}^2(Q) \equiv {\rm Av} \left[ \left| m^2(Q) \right| \right] \ ,
\end{equation}
where the average is taken over all scalar fields in the theory,
properly weighted by color and SU(2) factors. (This definition of
$\bar{m}^2(Q)$ is invariant under possible shifts from $D$-terms.)  We
then define the suppression, or crunching, factor
\begin{equation}
S \equiv \frac{\bar{m}^2(M_0)}{\bar{m}^2(m_W)} \ ,
\end{equation}
where the weak scale $m_W$ is taken to be 1 TeV.  In
Fig.~\ref{fig:mssmS} we plot $S$ in the $(m_N, h_G)$ plane.  For a
broad range of $h_t$ near its quasi-fixed point, we see that $S\sim
20$ is possible.  For smaller values of $h_t$, the scalar masses do
not approach their fixed point as quickly, and for larger values,
two-loop effects become important.  Indeed, the two-loop terms give
positive contributions to $\dot{Y}$, and therefore increase the Yukawa
suppression of scalar masses.  However, they also give positive
contributions directly to these masses.  (See Eq.~(\ref{mssmRGE}).)
The latter effects dominate, so the two-loop terms degrade the
hierarchy.  For $h_t$ larger than shown in Fig.~\ref{fig:mssmS}, the
two-loop effects become so large that perturbation theory cannot be
trusted.

The definition of $S$ is, of course, somewhat arbitrary, but its size
gives an indication of the reductions in the scalar masses that are
possible from radiative effects.  In particular, $S$ provides a rough
measure of the reduction of the scalar contribution to fine-tuning and
$\sqrt{S}$ is a measure of $\mheavy/\mlight$.  {}From
Fig.~\ref{fig:mssmS} we see that, if values of $\mlight$ of order 1
TeV are considered natural, values of $\mheavy$ up to $\sim 4 \tev$
are acceptable.  This is then the scale of the matter fields of the
first and second generation, which, given their small Yukawa
couplings, do not renormalize significantly.

A scale of 4 TeV does not completely solve the most stringent flavor
and $CP$ problems related to the $K$ system, but it does reduce them
significantly with respect to the typical case with squark masses of 1
TeV.  In addition, several of the other experimental difficulties
mentioned in Sec.~\ref{sec:introduction} are also resolved.  Detecting
scalars with such heavy masses will pose a serious challenge to future
colliders.

Finally, we note that the values of $S$ are largely independent of
$m_N$. For $m_N \alt 10^{13} \gev$, the hierarchy is already created
by the time the neutrino decouples, and so is insensitive to the
decoupling scale.  However, even for $m_N$ approaching $M_G$, a large
hierarchy is possible.  This may be understood as follows.  When $N$
decouples, the new evolution matrix $\boldN'$ is the upper $7\times 7$
block of $\boldN$ with $Y_n = 0$. The matrix $\boldN'$ has 3 positive
and 4 zero eigenvalues.  In terms of $\cos 3\theta = 17/64$, the 2
largest eigenvalues are $\frac{4} {3}(\cos \theta \pm \sqrt{3}
\sin\theta + 16) \approx 7.5, 5.6$; their corresponding eigenvectors
$\bolde'_1$ and $\bolde'_2$ are also quickly damped. If we decompose
the 7-dimensional truncation of $\bolde_1$ into these new
eigenvectors, $\bolde_1^{\rm tr} = (2,1,1,1,2,1,1) \equiv \sum_{i=1}^7
c'_i \bolde'_i$, we find that the decomposition is dominated by
$\bolde'_1$ and $\bolde'_2$ components: for the remaining components,
$|\sum_{i=3}^7 c'_i \bolde'_i | / | \bolde_1^{\rm tr} | = 0.20$.
Thus, the decoupling of $N$ perturbs the fixed point structure only
slightly --- when $N$ decouples, the scalar mass spectrum projects
mainly on to the new eigenvectors with largest eigenvalues, and the
rapid exponential damping continues.  Note that the special case $m_N
= M_G$ is equivalent to the case of an SU(5) GUT with universal Yukawa
couplings and minimal SU(5) particle content. The exact eigenvectors
for the SU(5) fixed point system are complicated.  However, the above
analysis shows that $\bolde_1^{\rm tr}$ is nearly a linear combination
of eigenvectors with large eigenvalues, so large hierarchies can be
generated from simple boundary conditions in the SU(5) case as well.

\section{GUTs above the Unification Scale}
\label{sec:guts}

An inverse hierarchy may also be generated in supergravity theories
through renormalization group evolution above the unification scale.
In this section we consider models in which scalar masses are
generated at the supergravity scale, which we take to be the reduced
Planck scale $M_\ast = 2.4 \times 10^{18} \gev$, and then evolved to
the GUT scale $M_G \simeq 2 \times 10^{16} \gev$.  Even though the
evolution interval is only two decades, we again find that substantial
hierarchies may be created for simple initial scalar spectra.  We will
consider two generic cases for illustration.

\subsection{SO(10)}

We begin by considering SO(10) models, that is, the class of models
with superpotential
\begin{equation}
W = h_t \psi \psi \bigH_2 + \ldots \ ,
\label{Wso10I}
\end{equation}
where the matter fields are contained in $\psi$ ({\boldmath $16$}) and
both up-type and down-type Higgs fields are contained in a single
multiplet: $\bigH_2$ ({\boldmath $10$}) $\supseteq (\bigH_u,
\bigH_d)$, where $\bigH_u$, $\bigH_d$ are {\boldmath $5$} and
{\boldmath $\bar{5}$} representations of the SU(5) subgroup,
respectively.  We allow arbitrary additional field content, subject
only to the constraint that additional superpotential couplings are
small relative to $h_t$.  Note that two Higgs models, with
superpotential $W = h_t \psi \psi \bigH_2 + h_b \psi \psi \bigH_1 +
\ldots$ are, for the purposes of our analysis, equivalent to a one
Higgs model in the limit $h_t = h_b$ (see Appendix).  To the extent
that the couplings are unified in this more general class of models,
this analysis also applies.

The RGEs for this class of models are given in the Appendix.  Note
that additional interactions and an extended Higgs sector are required
to break SO(10) and generate flavor structure.  In general, this leads
to a set of RGEs that are highly model-dependent.  Fortunately,
though, in the limit we consider, the RGEs simplify considerably.
First, we assume that all additional Yukawa couplings are small, so
their impact on the RGEs may be neglected.  Second, the presence of
additional fields with standard model quantum numbers modifies all RGE
terms with two or more powers of $\talpha$.  Fortunately, as argued in
Sec.~\ref{sec:mssm}, the two-loop terms of this form are highly
suppressed.  (We have checked that they are insignificant for
reasonable field content.)  Third, extra fields with standard model
quantum numbers also change the one-loop $\beta$-function coefficient
$b_{10}$.  To good approximation, this is the only important effect,
so the RGEs of these models may be studied simply by taking $b_{10}$
as an arbitrary free parameter.  We will consider the range $-3 \leq
b_{10} \leq 20$, for which $1/26 \alt \alpha(M_\ast) \alt 1/8$.

We now proceed as in Sec.~\ref{sec:mssm}.  The one-loop scalar mass
evolution is given by $\dot{\boldm}^2 = - Y_t {\boldN} \boldm^2$,
where
\begin{equation}
{\boldN}= \left[
\begin{array}{cc}
4 &  8 \\
5 & 10 \end{array} \right] \ ,
\end{equation}
and ${\boldm^2} = (m_{\bigH_2}^2, m_{\psi}^2)^T$.  The matrix $\boldN$
has eigenvector $\bolde_1 = (4,5)$ with eigenvalue 14, and another
eigenvector with eigenvalue zero.

\begin{figure}[t]
\postscript{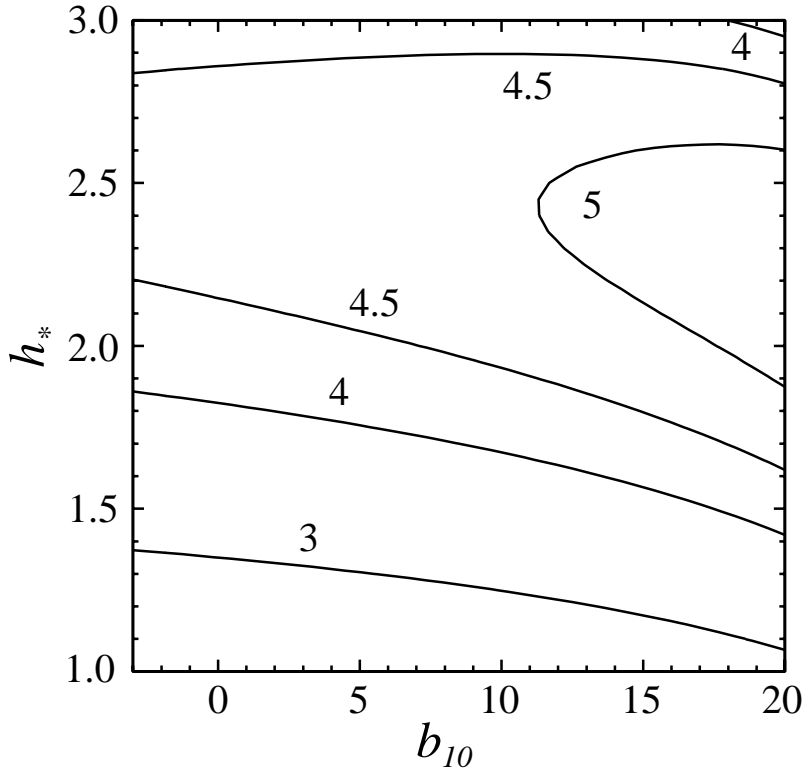}{0.65}
\caption{The suppression factor $S$ for SO(10) models evolving from
$M_\ast = 2.4 \times 10^{18}\gev$ to $M_G \simeq 2 \times 10^{16}
\gev$ with boundary conditions given by Eq.~(\ref{so10I}).  The
parameter $b_{10}$ is the one-loop $\beta$-function coefficient
parametrizing this class of models, and $h_\ast$ is the value of the
universal Yukawa coupling at the scale $M_\ast$.}
\label{fig:so10IS}
\end{figure}

We therefore consider what hierarchies are possible, starting with
initial boundary conditions
\begin{equation}
m_{\bigH_2}^2 = \frac{4}{5} m_{\psi}^2 \ .
\label{so10I}
\end{equation}
The suppression factor $S$ is plotted in Fig.~\ref{fig:so10IS}.
Several features are worthy of note.  First, $S$ is rather insensitive
to $b_{10}$, but is slightly improved for large particle content.  In
addition, even though the scalar masses are never pushed negative, $S$
is degraded for very large $h_\ast = h_t(M_\ast)$.  As above, this is
caused by two-loop effects.  Nevertheless, even for initial Yukawa
couplings $h_\ast \sim 2$, we see that $\sqrt{S} \sim 2$ is possible.
Scalars at 2 TeV again ameliorate a number of problems, and also
stretch the discovery limits of planned future colliders.

\subsection{SU(5)}

Finally, we consider the minimal SU(5) model and extensions.  The
minimal SU(5) model has superpotential
\begin{equation}
W = \frac{1}{4} h_t T T \bigH_u + \sqrt{2} h_b TF\bigH_d
+ \ldots\ ,
\end{equation}
where the quark and lepton fields are $T$ ({\boldmath $10$}) and $F$
({\boldmath $\bar{5}$}) and the Higgs fields are $\bigH_u$ ({\boldmath
$5$}) and $\bigH_d$ ({\boldmath $\bar{5}$}).

We assume that $h_t$ and $h_b$ are equal and large, and that all other
couplings are small relative to these.  Since we take $h_t$ and $h_b$
to be near their quasi-fixed points, the latter restriction is not too
severe.\footnote{Typically, the superpotential includes a term
$\lambda \bigH_u \Sigma \bigH_d$, where $\Sigma$ is the adjoint of
SU(5).  The requirement that colored Higgses be sufficiently heavy to
prevent proton decay then implies $\lambda \sim g$.  However, in our
scenarios, the first and second generation squarks are very massive,
and so such constraints are relaxed by either a factor of
$(\mlight/\mheavy)^{4}$ and an additional mixing angle or by a product
of two additional mixing angles.}  Note that we must take $h_t$ and
$h_b$ to be large in order to suppress $m_{H_u}$ {\it and} $m_{H_d}$.
This was an automatic and attractive feature of the minimal SO(10)
models.

The RGEs are given in the Appendix.  In the minimal model, the
$\beta$-function coefficient is $b_5 = -3$; it is larger in extensions
of the minimal model, and so, as before, we take it to be a free
parameter.

\begin{figure}[t]
\postscript{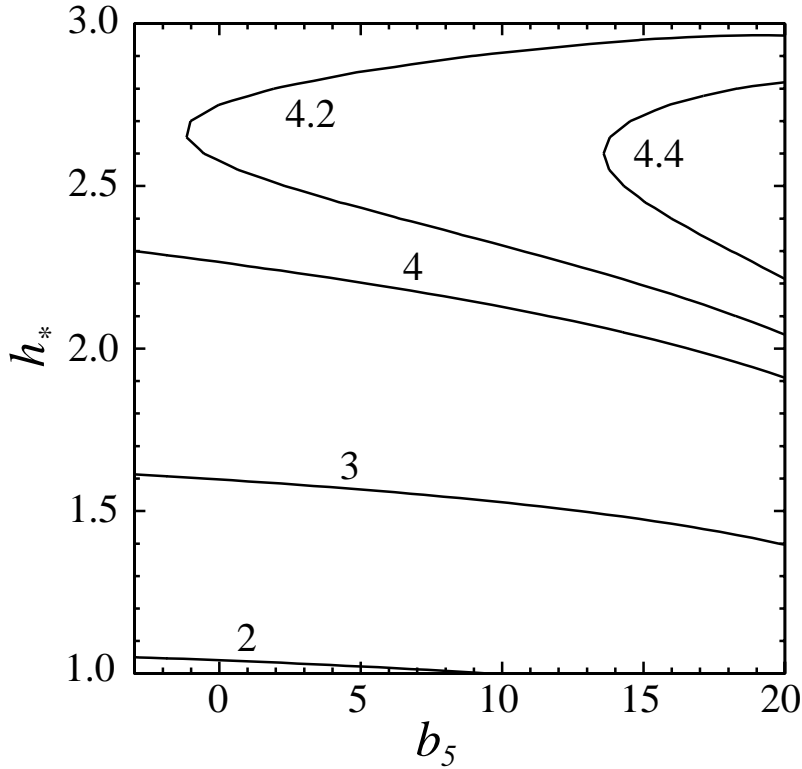}{0.65}
\caption{The suppression factor $S$, as in Fig.~\ref{fig:so10IS}, but
for SU(5) theories with the boundary conditions of Eq.~(\ref{su5}).}
\label{fig:su5S}
\end{figure}

The evolution matrix is
\begin{equation}
{\boldN}= \left[
\begin{array}{cccc}
3 &  6 & 0  & 0  \\
3 &  8 & 2  & 2  \\
0 &  4 & 4  & 4  \\
0 &  4 & 4  & 4  \end{array} \right]
\end{equation}
in the basis ${\boldm^2} = (m_{\bigH_u}^2, m_{T}^2, m_{F}^2,
m_{\bigH_d}^2)^T$, where $Y=Y_t=Y_b$.  The leading eigenvector is
$\bolde_1 = (3,5,4,4)$, with eigenvalue 13.  Suppression factors, with
initial conditions
\begin{equation}
\frac{3}{4} m_{\bigH_u}^2 = \frac{5}{4} m_{T}^2 = m_{F}^2 =
m_{\bigH_d}^2\ ,
\label{su5}
\end{equation}
are plotted in Fig.~\ref{fig:su5S}.  We find results similar to the
SO(10) case.  First and second generation scalar masses as large as 2
TeV are allowed.  As before, they significantly reduce the stringency
of the experimental constraints on scalar degeneracy and
$CP$-violating phases.

\section{Conclusions}
\label{sec:conclusions}

In this study we have examined the possibility that soft supersymmetry
breaking scalar masses are of order $\mheavy$ at some high scale.  The
third generation scalar masses are driven to $\mlight \alt \tev$ in
the infrared by their large Yukawa couplings, while first and second
generation scalar masses remain at $\mheavy$.  This inverse hierarchy
mechanism offers an appealing and natural way to satisfy the strong
experimental constraints on first and second generation scalars in
supersymmetric theories.

The third generation scalar masses are determined by infrared fixed
points, so their precise values are insensitive to the initial
conditions at, say, $M_G$ (or to the exact value of $M_G$, for that
matter).  We have shown that with suitable assumptions and
approximations, the scalar fixed points may be extracted from the RGEs
analytically.  We justified the validity of these approximations by
more precise numerical calculations.

For evolution below the GUT scale, we find that hierarchies of
$\mheavy/\mlight \sim 4$ may be created.  The necessary boundary
conditions are those of Eq.~(\ref{mssmbc}), with the Higgs squared
masses double those of the squarks and sleptons.  The simplicity of
this boundary condition is in some sense a measure of the plausibility
of the scenario.  It is remarkable that such a simple boundary
condition, perhaps the simplest imaginable next to complete scalar
universality, allows one to move the entire scalar mass scale up to 4
TeV without sacrificing naturalness.  Note that for simplicity we have
focussed on only the eigenvectors with largest eigenvalue.  However,
other eigenvectors with non-zero eigenvalue are also crunched, and so
initial conditions in these directions may also be acceptable.  It
would be interesting to find theories that naturally generate such
boundary conditions and to investigate the fixed point structure of
other models.

The scale $\mheavy \sim 4 \tev$ eliminates many dangerous
supersymmetric effects; for example, supersymmetric contributions to
EDMs, which in these scenarios scale like $\mlight^{2}/\mheavy^{4}$,
may be easily below the experimental bounds even for ${\cal O}(1)$
phases. However, these hierarchies are not sufficient to completely
satisfy the most stringent experimental bounds by themselves.  For
example, for $\mheavy \sim 4 \tev$, the $K-\bar{K}$ mass difference
still requires $m_{12}^2 / \mheavy^2 \alt 0.1$, where $m_{12}^2$ is
the off-diagonal entry of the squark mass matrix mixing the first and
second generations.  Nevertheless, such constraints are significantly
weaker than in typical scenarios.

For evolution above the GUT scale, the brief evolution interval makes
it impossible to generate such large hierarchies.  However,
$\mheavy/\mlight \sim 2$ is possible, and constraints on scalar
degeneracy and $CP$-violating phases are still significantly weakened
relative to the standard scenario in which all scalars are below 1
TeV.  Note that this is opposite to the typical situation is unified
models, where ${\cal O}(1)$ generational splitting from
unification-era evolution \cite{npap} can lead to new contributions to
flavor \cite{bh,gg} and $CP$ violation \cite{dh}.  (A similar
situation occurs in the presence of a right-handed neutrino with a
large Yukawa coupling \cite{rhn}.)

Throughout this analysis, we have neglected off-diagonal scalar
squared masses.  For the squark masses, constraints on $m_{12}$ have
already been noted above.  The off-diagonal elements $m_{13}$ and
$m_{23}$ are also constrained: the requirement that the squark mass
matrix be tachyon-free implies $m_{13}^2, m_{23}^2 \alt \mlight
\mheavy$.  The off-diagonal masses are not driven to $\mlight$ and
are, in fact, essentially RGE-invariant; the origins of their slight
suppressions must therefore lie elsewhere.  Depending on the extent of
these suppressions, some interesting signals, for example, in the $B$
system, may be observable in the near future~\cite{bphys}.

Finally, we draw some implications for high energy colliders.  In
these scenarios, as in all scenarios with inverted scalar hierarchies,
scalars with mass $\mheavy$ are beyond the reach of proposed
colliders, such as the LHC and NLC, and can only be detected through
their non-decoupling from super-oblique corrections~\cite{so}.  A goal
of supersymmetric collider studies is to precisely measure
superparticle properties, and thereby determine the supersymmetric
parameters at the weak scale.  It is then hoped that hints about
physics at even smaller distances can be found by evolving these
parameters to very high energies.  The fixed point analysis shows that
in certain directions of parameter space, corresponding to
eigenvectors of the scalar mass evolution matrix with large
eigenvalues, even large masses at the high scale are exponentially
suppressed when they are evolved to the weak scale.  Accurate
determination of high scale boundary conditions in these directions
therefore requires extremely precise measurements.

\acknowledgements

It is a pleasure to thank Takeo Moroi for helpful discussions.  JAB is
supported by the National Science Foundation, grant NSF--PHY--9404057,
and by the Monell Foundation.  JLF is supported by the Department of
Energy under contract DE--FG02--90ER40542 and through the generosity
of Frank and Peggy Taplin.  The work of NP is supported by the
Department of Energy under contract DE--FG02--96ER40559.

\appendix

\section*{}

In this appendix, we present the renormalization group equations for
SU(5) and SO(10) GUTs above and below the GUT scale.  As explained in
Sec.~\ref{sec:mssm}, gaugino masses and $A$ terms are negligible in
the scenarios we are considering.  The relevant parameters are then
the gauge couplings $g$, Yukawa couplings $h$, and scalar squared
masses $m^2$.  The RGEs are conveniently expressed in terms of the
variables
\begin{eqnarray}
t &\equiv& \ln \left(M_0^2/Q^2\right) \\
\talpha &\equiv& \frac{g^2}{16 \pi^2} = \frac{\alpha}{4 \pi} \\
Y &\equiv& \frac{h^2}{16 \pi^2} \\
X &\equiv& m_X^2 \label{notation} \ ,
\end{eqnarray}
where, as indicated in Eq.~(\ref{notation}), for notational simplicity
we denote a scalar field's squared mass by the field itself.

Schematically, the two-loop RGEs then have the form
\begin{eqnarray}
\dot{\talpha} &=& \underline{b\talpha^2} + \underline{\talpha^2 Y}
+ \talpha^3 \nonumber \\
\dot{Y} &=& Y \left[\underline{-Y} + \underline{\talpha}
+ \underline{Y^2} - \talpha Y - \talpha^2\right] \nonumber \\
\dot{X} &=& X \left[ \underline{-Y} + \underline{Y^2} - \talpha Y -
\talpha^2 \right] \ ,
\end{eqnarray}
where $\dot{(\ )} \equiv \frac{d}{dt}$.  The constant $b$ is the
one-loop $\beta$-function coefficient; the signs of all other terms
are determined as indicated.  Most of the two-loop terms are
insignificant, as described in Sec.~\ref{sec:mssm}.  Below, we present
all one-loop terms and the dominant two-loop terms underlined above.

\subsection{MSSM with Right-handed Neutrino}

The superpotential for the MSSM with a right-handed neutrino
supermultiplet $N$ is
\begin{equation}
W = h_t H_u Q U + h_b H_d Q D + h_{\tau} H_d L E + h_n H_u L N \ .
\end{equation}

The RGEs for the three (GUT normalized) gauge couplings are
\begin{eqnarray}
\dot{\talpha}_3 &=& 3 \talpha_3^2+\talpha_3^2 (4Y_t+4Y_b)
\nonumber \\
\dot{\talpha}_2&=&-\talpha_2^2+\talpha_2^2 (6Y_t+6Y_b+2Y_\tau+2Y_n)
\nonumber \\
\dot{\talpha}_1&=&-\frac{33}{5}\talpha_1^2+\talpha_1^2
\left(\frac{26}{5}Y_t+\frac{14}{5}Y_b+\frac{18}{5}Y_\tau+\frac{6}{5}Y_n
\right)\ .
\end{eqnarray}

The Yukawa coupling RGEs are
\begin{eqnarray}
\dot{Y}_t &=& Y_t\ [-6Y_t-Y_b-Y_n+\frac{16}{3}\talpha_3+3\talpha_2
+\frac{13}{15}\talpha_1 \nonumber \\
&& \quad + 22Y_t^2+5Y_t Y_b +3 Y_t Y_n + 5Y_b^2 + Y_b
Y_\tau + Y_\tau Y_n + 3 Y_n^2 ] \nonumber \\
\dot{Y}_b&=&Y_b\ [-Y_t-6Y_b-Y_\tau+\frac{16}{3}\talpha_3+3\talpha_2
+\frac{7}{15}\talpha_1 \nonumber \\
&& \quad + 5Y_t^2+5Y_t Y_b + Y_t Y_n + 22Y_b^2 + 3 Y_b
Y_\tau + 3 Y_\tau^2 + Y_\tau Y_n] \nonumber \\
\dot{Y}_\tau&=&Y_\tau \ [-3Y_b-4Y_\tau-Y_n+3\talpha_2
+\frac{9}{5}\talpha_1 \nonumber \nonumber \\
&& \quad + 3Y_t Y_b +3 Y_t Y_n + 9Y_b^2 + 9Y_b
Y_\tau + 10 Y_\tau^2 + 3 Y_\tau Y_n + 3 Y_n^2] \nonumber \\
\dot{Y}_n&=&Y_n\ [-3Y_t-Y_\tau-4Y_n+3\talpha_2 +\frac{3}{5}\talpha_1
\nonumber \\
&& \quad +9Y_t^2+3Y_t Y_b +9 Y_t Y_n + 3Y_b
Y_\tau + 3Y_\tau^2 +3Y_\tau Y_n + 10 Y_n^2] \ .
\end{eqnarray}

Finally, assuming that the hypercharge trace ${\rm tr} [Y' m^2]$
vanishes, the scalar squared masses evolve as
\begin{eqnarray}
\dot{Q}&=& -Y_t s_1 - Y_b s_2 + 10 Y_t^2 s_1 + 10 Y_b^2 s_2
+ Y_t Y_n (s_1 + s_4) + Y_b Y_\tau (s_2 + s_3) \nonumber \\
\dot{U}&=& -2Y_t s_1 + 16 Y_t^2 s_1 + 2 Y_t Y_b (s_1 + s_2)
+ 2 Y_t Y_n (s_1 + s_4) \nonumber \\
\dot{D}&=& - 2 Y_b s_2 + 16 Y_b^2 s_2 + 2 Y_t Y_b (s_1+s_2)
+ 2 Y_b Y_\tau (s_2 + s_3) \nonumber \\
\dot{L}&=& -Y_\tau s_3 - Y_n s_4 + 6 Y_\tau^2 s_3 + 6 Y_n^2 s_4
+ 3 Y_t Y_n (s_1 + s_4) + 3 Y_b Y_\tau (s_2 + s_3) \nonumber \\
\dot{E}&=& -2 Y_\tau s_3 + 8 Y_\tau^2 s_3 + 2 Y_\tau Y_n (s_3 + s_4)
+ 6 Y_b Y_\tau (s_2 + s_3) \nonumber \\
\dot{N}&=& - 2 Y_n s_4 + 8 Y_n^2 s_4
+ 2 Y_\tau Y_n (s_3 + s_4) + 6 Y_t Y_n (s_1 + s_4) \nonumber \\
\dot{H}_u&=& -3 Y_t s_1 - Y_n s_4 + 18 Y_t^2 s_1+3Y_t Y_b (s_1+s_2)
+ 6 Y_n^2 s_4 + Y_\tau Y_n (s_3 + s_4) \nonumber \\
\dot{H}_d&=& -3Y_b s_2 - Y_\tau s_3 + 18 Y_b^2 s_2
+ 3 Y_t Y_b (s_1 + s_2) + 6 Y_\tau^2 s_3 + Y_\tau Y_n (s_3 + s_4) \ ,
\end{eqnarray}
where
\begin{eqnarray}
s_1 &\equiv& H_u + Q + U \nonumber \\
s_2 &\equiv& H_d + Q + D \nonumber \\
s_3 &\equiv& H_d + L + E \nonumber \\
s_4 &\equiv& H_u + L + N \ .
\end{eqnarray}

These RGEs are valid above the right-handed neutrino mass scale $m_N$.
The RGEs for the MSSM with minimal field content may be obtained, of
course, by setting $Y_n = 0$.

\subsection{SO(10) above the GUT scale}

In the text, we consider a class of minimal SO(10) models with only
one large Yukawa coupling.  Here, for completeness, we present RGEs
for the more general class of models specified by the superpotential
\begin{equation}
W = h_t \psi \psi \bigH_2 + h_b \psi \psi \bigH_1 + \ldots \ ,
\end{equation}
where the quark and lepton fields are contained in $\psi$ ({\boldmath
$16$}), and $\bigH_2$ ({\boldmath $10$}) and $\bigH_1$ ({\boldmath
$10$}) are Higgs fields.  In addition to the usual scalar mass terms,
we also include the off-diagonal mass $m_{\bigH_{12}}^2
(\bigH_1^{\dagger} \bigH_2 + \bigH_2^{\dagger} \bigH_1)$, which is
allowed by all symmetries and is generated even if it is initially set
to zero.  The RGEs presented here correct certain one-loop terms given
in Ref.~\cite{bh}, where the effects of this term were omitted.

We consider models with arbitrary additional field content, but with
interactions such that the dominant Yukawa couplings are $h_t$ and
$h_b$.  The RGEs are then
\begin{equation}
\dot{\talpha} = - b_{10} \talpha^2+\talpha^2 (28Y_t+28Y_b) \ ,
\end{equation}
\begin{eqnarray}
\dot{Y}_t &=& Y_t\ [-14Y_t-14Y_b+\frac{63}{2}\talpha
+ 130Y_t^2+ 260 Y_t Y_b +130 Y_b^2 ] \nonumber \\
\dot{Y}_b &=& Y_b\ [-14Y_t-14Y_b+\frac{63}{2}\talpha
+ 130Y_t^2+ 260 Y_t Y_b +130 Y_b^2 ] \ ,
\end{eqnarray}
and
\begin{eqnarray}
\dot{\psi}&=& -5 Y_t u_1 - 5 Y_b u_2 - 10 Y_{tb} \bigH_{12}
\nonumber \\
&& + 90 Y_t^2 u_1 + 90 Y_t Y_b (u_1+u_2) + 90 Y_b^2 u_2
+ 180 (Y_t + Y_b) Y_{tb} \bigH_{12} \nonumber \\
\dot{\bigH}_2&=& -4 Y_t u_1 - 4 Y_{tb} \bigH_{12} \nonumber \\
&& + 80 Y_t^2 u_1 + 40 Y_t Y_b (u_1+u_2)
+ 120 Y_t Y_{tb} \bigH_{12} + 40 Y_b Y_{tb} \bigH_{12}  \nonumber \\
\dot{\bigH}_1&=& -4 Y_b u_2 - 4 Y_{tb} \bigH_{12} \nonumber \\
&& + 80 Y_b^2 u_2 + 40 Y_t Y_b (u_1+u_2)
+ 120 Y_b Y_{tb} \bigH_{12} + 40 Y_t Y_{tb} \bigH_{12}  \nonumber \\
\dot{\bigH}_{12}&=& -2 Y_{tb} (u_1 + u_2) - 2 (Y_t + Y_b) \bigH_{12}
\nonumber \\
&&+ 20 Y_t Y_{tb} (3 u_1 + u_2) + 20 Y_b Y_{tb} (3 u_2 + u_1)
+ (20 Y_t^2 + 120 Y_t Y_b + 20 Y_b^2) \bigH_{12} \ ,
\end{eqnarray}
where
\begin{eqnarray}
u_1 &\equiv& 2\psi + \bigH_2 \nonumber\\
u_2 &\equiv& 2\psi + \bigH_1 \nonumber\\
\bigH_{12} &\equiv& m_{\bigH_{12}}^2 \nonumber\\
Y_{tb}^2 &\equiv& Y_t Y_b \ ,
\end{eqnarray}
and
\begin{equation}
b_{10} = -16 + S(R) \ .
\end{equation}
Common representations of SO(10) and their Dynkin indices are
{\boldmath $10$} (1), {\boldmath $16$} (2), {\boldmath $45$} (8),
{\boldmath $54$} (12), {\boldmath $120$} (28), and {\boldmath $126$}
(35).

In minimal SO(10) models, both up-type and down-type Higgs multiplets
are contained in one SO(10) multiplet: $\bigH_2$ ({\boldmath $10$})
$\supseteq (\bigH_u, \bigH_d)$, where $\bigH_u$ and $\bigH_d$ are
{\boldmath $5$} and {\boldmath $\bar{5}$} representations of the SU(5)
subgroup, respectively. The RGEs for these models are obtained by
setting $h_b = \bigH_1 = \bigH_{12} = 0$ in the equations above.  The
RGEs above, however, are also applicable to models in which the Higgs
fields are contained in two separate SO(10) representations, with
$\bigH_2$ ({\boldmath $10$}) $\supseteq (\bigH_u, \bar{\bigH_u})$ and
$\bigH_1$ ({\boldmath $10$}) $\supseteq (\bar{\bigH}_d, \bigH_d)$.

A useful consistency check follows from noting that, for $h_t = h_b =
h/\sqrt{2}$ and $\bigH_1 = \bigH_2 = \bigH_{12} = \bigH/2$, the two
Higgs model reduces to a minimal model with Higgs field $(\bigH_2 +
\bigH_1)/\sqrt{2}$, and indeed, the resulting RGEs are identical to
those for minimal models.

\subsection{SU(5) above the GUT scale}

We consider also the minimal SU(5) model and possible extensions.  The
minimal SU(5) model has superpotential
\begin{equation}
W = \frac{1}{4} h_t T T \bigH_u + \sqrt{2} h_b TF\bigH_d
+ \mu_H \bigH_u \bigH_d + \lambda \bigH_u \Sigma \bigH_d
+ \mu_\Sigma \tr \Sigma^2 + \frac{1}{6} \lambda' \tr \Sigma^3 \ ,
\end{equation}
where the quark and lepton fields are $T$ ({\boldmath $10$}) and $F$
({\boldmath $\bar{5}$}) and the Higgs fields are $\bigH_u$ ({\boldmath
$5$}), $\bigH_d$ ({\boldmath $\bar{5}$}), and $\Sigma$ ({\boldmath
$24$}).

The Yukawa couplings $h_t$ and $h_b$ are large and near their fixed
point.  We assume that all other couplings, such as $\lambda$,
$\lambda'$, $\mu_H$, and $\mu_\Sigma$, are small relative to these.
The RGEs are then
\begin{equation}
\dot{\talpha}=-b_5\talpha^2+\talpha^2 (12Y_t+14Y_b) \label{alpha5}\ ,
\end{equation}
\begin{eqnarray}
\dot{Y}_t &=& Y_t\ [-9Y_t-4Y_b+\frac{96}{5}\talpha
+ 54Y_t^2+ 24 Y_t Y_b +32 Y_b^2 ] \nonumber \\
\dot{Y}_b &=& Y_b\ [-3Y_t-10Y_b+\frac{84}{5}\talpha
+ 18Y_t^2+ 30 Y_t Y_b +64 Y_b^2 ] \ ,
\end{eqnarray}
and
\begin{eqnarray}
\dot{T}&=& -3 Y_t t_1 - 2 Y_b t_2 + 36 Y_t^2 t_1
+ 6 Y_t Y_b (t_1+t_2) + 32 Y_b^2 t_2 \nonumber \\
\dot{F}  &=& -4 Y_b t_2 + 12 Y_t Y_b (t_1+t_2) + 48 Y_b^2 t_2
\nonumber \\
\dot{\bigH}_u&=& -3 Y_t t_1 + 36 Y_t^2 t_1 + 12 Y_t Y_b (t_1+t_2)
\nonumber \\
\dot{\bigH}_d&=& -4 Y_b t_2 + 12 Y_t Y_b (t_1+t_2) + 48 Y_b^2 t_2 \ ,
\end{eqnarray}
where
\begin{eqnarray}
t_1 &\equiv& 2T + \bigH_u \nonumber \\
t_2 &\equiv& T + F + \bigH_d \ .
\end{eqnarray}

In the minimal model, $b_5 = -3$ in Eq.~(\ref{alpha5}).  For
extensions of the minimal model, the above equations apply (again,
assuming no large Yukawas), with the modification
\begin{equation}
b_5 = -3+ S(R) \ ,
\end{equation}
where $S(R)$ is the sum of the Dynkin indices of the additional
fields.  For convenience, a few typical SU(5) representations and
their Dynkin indices are listed here: {\boldmath $5$} ($\frac{1}{2}$),
{\boldmath $10$} ($\frac{3}{2}$), {\boldmath $15$} ($\frac{7}{2}$),
{\boldmath $24$} (5), {\boldmath $45$} (12), {\boldmath $50$}
($\frac{35}{2}$), and {\boldmath $75$} (25).

\end{document}